\documentclass{aastex631}
\usepackage{subfigure}
\usepackage{amsmath}
\usepackage{bm}
\usepackage{amsmath,amsthm,amssymb,amsfonts}
\usepackage{multirow}
\usepackage{verbatim}
\usepackage{threeparttable}
\usepackage{tabularx}
\usepackage{booktabs}
\usepackage{xcolor}
 
\usepackage{textcomp}
\usepackage{hyperref}
\usepackage[misc]{ifsym} 

\graphicspath{{./}{figures/}}

\begin{document}

\title{Multibeam SETI Observations toward Nearby M dwarfs with FAST}

\author[0000-0003-3977-4276]{Xiao-Hang Luan}
\affiliation{Institute for Frontiers in Astronomy and Astrophysics, Beijing Normal University, Beijing 102206, People's Republic of China}
\affiliation{School of Physics and Astronomy, Beijing Normal University, Beijing 100875, People's Republic of China; \url{tjzhang@bnu.edu.cn}}

\author{Bo-Lun Huang}
\affiliation{Institute for Frontiers in Astronomy and Astrophysics, Beijing Normal University, Beijing 102206, People's Republic of China}
\affiliation{School of Physics and Astronomy, Beijing Normal University, Beijing 100875, People's Republic of China; \url{tjzhang@bnu.edu.cn}}

\author[0000-0002-4683-5500]{Zhen-Zhao Tao}
\affiliation{Institute for Astronomical Science, Dezhou University, Dezhou 253023, People's Republic of China}

\author{Yan Cui \href{mailto:yancui@sdu.edu.cn}{\textrm{\Letter}}}
\affiliation{School of Law, Gansu University of Political Science and Law, Lanzhou 730070, People's Republic of China; \url{yancui@sdu.edu.cn}}

\author[0000-0002-3363-9965]{Tong-Jie Zhang\href{mailto:tjzhang@bnu.edu.cn}{\textrm{\Letter}}}
\affiliation{Institute for Frontiers in Astronomy and Astrophysics, Beijing Normal University, Beijing 102206, People's Republic of China}
\affiliation{School of Physics and Astronomy, Beijing Normal University, Beijing 100875, People's Republic of China; \url{tjzhang@bnu.edu.cn}}
\affiliation{Institute for Astronomical Science, Dezhou University, Dezhou 253023, People's Republic of China}

\author{Pei Wang}
\affiliation{Institute for Frontiers in Astronomy and Astrophysics, Beijing Normal University, Beijing 102206, People's Republic of China}
\affiliation{National Astronomical Observatories, Chinese Academy of Sciences, Beijing 100101, People's Republic of China}


\begin{abstract}
The search for extraterrestrial intelligence (SETI) targeted searches aim to observe specific areas and objects to find possible technosignatures. Many SETI researches have focused on nearby stars and their planets in recent years. In this paper, we report a targeted SETI observations using the most sensitive L-band Five-hundred-meter Aperture Spherical radio Telescope (FAST) toward three nearby M dwarfs, all of which have been discovered exoplanet candidates. The minimum equivalent isotropic radiant power of the lower limit from the three sources we can detect is $6.19 \times 10^{8}$ W, which is well within the reach of current human technology. Applying the multibeam coincidence matching (MBCM) blind search mode, we search for narrowband drifting signals across $1.05-1.45$ GHz in each of the two orthogonal linear polarization directions. An unusual signal at 1312.50 MHz detected from the observation toward AD Leo originally piqued our interest. However, we finally 
eliminate the possibility of an extraterrestrial origin based on much evidence, such as the polarization, frequency, and beam coverage characteristics.

\end{abstract}


\section{Introduction} \label{sec:intro}

The search for extraterrestrial intelligence (SETI) seeks to address one of the most profound questions: is there any life beyond Earth? In contrast to methods that search for biosignatures (e.g., \citet{2014Sci...343..171R,2015Sci...347..415W}), SETI focuses on technosignatures, which are signals intentionally or unintentionally emitted by intelligent civilizations. Since the 1960s \citep{1961PhT....14d..40D}, SETI has predominantly utilized radio observations due to the effectiveness of radio electromagnetic signal propagation through interstellar space.

Radio SETI observations are usually conducted in two modes: sky surveys and targeted searches \citep{2001ARA&A..39..511T}. Sky surveys, such as SERENDIP, SETI@home \citep{2001SPIE.4273..104W}, and FAST SETI backend \citep{2020ApJ...891..174Z,2023AJ....166..146W}, perform commensal observations over extensive areas of the sky. On the other hand, targeted searches focus on high-potential areas and objects, such as exoplanet systems \citep{2013ApJ...767...94S,2016ApJ...827L..22T,2016AJ....152..181H,2018ApJ...869...66H,2019AJ....157..122P,2021AJ....161..286T,2022AJ....164..160T,2023AJ....165..132L}, nearby stars \citep{2017ApJ...849..104E,2020AJ....159...86P,2020AJ....160...29S,2021NatAs...5.1148S,2021AJ....161...55M,2023AJ....166..190T,2023NatAs...7..492M}, nearby galaxies \citep{2017AJ....153..110G}, galactic center \citep{2020PASA...37...35T,2021AJ....162...33G}, and galactic anticenter \citep{2018ApJ...856...31T,2020PASA...37...35T}. Most radio SETI studies concentrate on detecting narrowband signals ($\sim$ Hz) because these signals are commonly used in human communications and are easily distinguishable from natural astrophysical sources \citep{1987MNRAS.225..491C}. Despite extensive efforts, no definitive evidence of extraterrestrial technosignatures has been confirmed.

As the largest single-aperture radio telescope, the sensitivity of FAST \citep{2006ScChG..49..129N,2011IJMPD..20..989N,2016RaSc...51.1060L,2020RAA....20...64J} offers significant opportunities for SETI research \citep{2020RAA....20...78L,2021RAA....21..178C}. To identify and mitigate radio frequency interference (RFI) from human technology, two primary observation and search strategies are employed with the FAST 19-beam receiver: the multibeam coincidence matching (MBCM) strategy \citep{2022AJ....164..160T,2023AJ....166..190T,2023AJ....165..132L} and the multibeam point-source scanning (MBPS) strategy \citep{2023AJ....166..245H}. The MBCM strategy involves simultaneous observations with multiple beams, comparing their difference to identify RFI. It offers advantages such as higher time efficiency and more reference positions compared to the traditional on-off strategy \citep{2013ApJ...767...94S,2017ApJ...849..104E,2019AJ....157..122P,2020AJ....159...86P,2020AJ....160...29S,2021AJ....162...33G,2021AJ....161..286T,2021NatAs...5.1148S}. The MBCM strategy can be executed in two modes: targeted search mode \citep{2022AJ....164..160T,2023AJ....166..190T} and blind search mode \citep{2023AJ....165..132L}, where both modes are for targeted SETI observations. Applying the MBCM strategy, we searched for narrowband signals toward exoplanet systems in previous studies \citep{2022AJ....164..160T,2023AJ....165..132L}. As a supplement to previous SETI observations, we select nearby stars as observation targets this time.

M dwarfs, the most abundant stars in the Milky Way, are cool main-sequence low-mass stars. They constitute 75\% of the stars in the solar neighborhood \citep{2002AJ....124.2721R,2006AJ....132.2360H}. M dwarfs are highly active stars, exhibiting more stellar activity than the Sun, often showing strong magnetic fields \citep{1985ApJ...299L..47S,2009ApJ...692..538R,2017NatAs...1E.184S,2021A&ARv..29....1K}, spots, flares, and other brightness inhomogeneities. In recent years, one of the primary goals of exoplanet searches has been to identify habitable planets around M dwarfs. Due to their low mass and luminosity, M dwarfs are ideal targets for detecting Earth-like planets using the radial velocity (RV) method \citep{2012A&A...541A...9G}. Observationally, the likelihood of finding Earth-like planets in the habitable zone (HZ) increases as the mass of the host star decreases. According to \citet{2015ApJ...807...45D}, M dwarfs have at least 2.5 $\pm$ 0.2 planets per star. Moreover, the occurrence rate of Earth-sized planets is estimated at $0.16_{-0.07}^{+0.17}$ per M dwarf, and for super-Earths, it is $0.12_{-0.05}^{+0.10}$ per M dwarf within a conservatively defined HZ. Adopting the broader insolation limits of recent Venus and early Mars results in a higher estimate of $0.24_{-0.08}^{+0.18}$ Earth-sized planets and $0.21_{-0.06}^{+0.11}$ super-Earths per M dwarf HZ \citep{2015ApJ...807...45D}. However, the high stellar activity of M dwarfs may impact the structure and temperature of exoplanet atmospheres, which in turn affects the size of the HZ \citep{2007AsBio...7..185L} and could negatively impact the survival and development of nearby life \citep{2017ApJ...841..124V}.

In this paper, we conduct a series of SETI observations on three nearby M dwarfs using the MBCM blind search mode. The rest of this paper is organized as follows. We describe the targets and the strategies in Section \ref{sec:observe}, then discuss the data analysis in Section \ref{sec:analysis}. The results of the signal search and RFI removal are presented in Section \ref{sec:results}. In Section \ref{sec:discussions}, we discuss the sensitivity of our observations and the power variation in two polarizations. We finally conclude this work in Section \ref{sec:conclusions}.

\section{Observations} \label{sec:observe}

\subsection{Targets} \label{subsec:targets}

Due to the abundance of M dwarfs in the Milky Way and the high possibility of each star having planets, we select three typical M dwarfs, Wolf 359, AD Leo, and TVLM 513-46546, all of which have been discovered exoplanet candidates and locate within 10 pc. These three targets are observed for the first time in the FAST SETI research.

Wolf 359 is the fifth closest star to the Sun and has relatively strong flaring activities, which has been monitored from radio to X-ray bands \citep{2007A&A...468..221F,2008A&A...487..293F,2010A&A...511A..83F,2010A&A...514A..94L}. The RV measurements suggested that Wolf 359 is a planetary host \citep{2019arXiv190604644T}, although its HZ has not yet been thoroughly explored. AD Leo is a flare star whose intense flaring activities have been detected in radio, visible light, extreme-ultraviolet, and X-ray bands \citep{1986ApJ...305..363L,1989A&A...220L...5G,1990ApJ...353..265B,1997A&A...321..841A,1997ApJ...491..910C,2003ApJ...597..535H,2005A&A...435.1073R,2008ApJ...674.1078O,2020A&A...637A..13M,2023ApJ...953...65Z}. \citet{2018AJ....155..192T} claimed the possible existence of a hot giant planet. However, this claim has been questioned by some subsequent studies \citep{2020A&A...638A...5C,2022A&A...666A.143K}. TVLM 513-46546 is an ultracool dwarf exhibiting flaring activity, which is most pronounced at radio wavelengths \citep{2006ApJ...653..690H,2013ApJ...779..101H,2014ApJ...788...23W}. \citet{2020AJ....160...97C} announced the discovery of a Saturn-like planet around this star. The detailed information of the three targets are shown in Table \ref{table:targets}.

On June 24, 2023, we observed these three nearby M dwarfs successively with FAST (the start MJDs are 60119.290972, 60119.320139, and 60119.513194 for AD Leo, Wolf 359, and TVLM 513-46546, respectively), each lasting 30 minutes.

\begin{table*}[!ht]
	\caption{Targets}
	\centering
	\begin{threeparttable}
	\begin{tabularx}{\textwidth}{lcccccc}
		\hline \hline Star Name & R.A. (J2000)\tnote{a} & Decl. (J2000)\tnote{a} & Distance\tnote{a} (pc) & Spectral Type & Exoplanet Candidates\\
		\hline Wolf 359 & 10:56:28.92 & +07:00:53.00 & 2.41 & M6 V\tnote{b} & Wolf 359 b, Wolf 359 c\tnote{e}\\
		AD Leo & 10:19:36.28 & +19:52:12.01 & 4.97 & M4.5 V\tnote{c} & AD Leo b\tnote{f}\\
		TVLM 513-46546 & 15:01:08.19 & +22:50:02.14 & 10.73 & M9 V\tnote{d} & TVLM 513 b\tnote{g}\\
		\hline
	\end{tabularx}
        \begin{tablenotes}[flushleft]
      	 \footnotesize
      	 \item{\bf Notes.}
      	 \item[a] From \citet{2023A&A...674A...1G}.
      	 \item[b] From \citet{2019AJ....157...63K}.
      	 \item[c] From \citet{2013A&A...552A.103R}.
      	 \item[d] From \citet{2020AJ....160...97C}.
      	 \item[e] From \citet{2019arXiv190604644T}. However, Wolf 359 c was refuted by \citet{2021A&A...652A..28L}.
      	 \item[f] From \citet{2018AJ....155..192T}. However, it has been questioned by \citet{2020A&A...638A...5C} and \citet{2022A&A...666A.143K}.
      	 \item[g] From \citet{2020AJ....160...97C}.
        \end{tablenotes}
  \end{threeparttable}
	\label{table:targets}
\end{table*}

\begin{figure*}[!htp]
	
	\centering
	\subfigure[]{
		\includegraphics[width=0.29\linewidth]{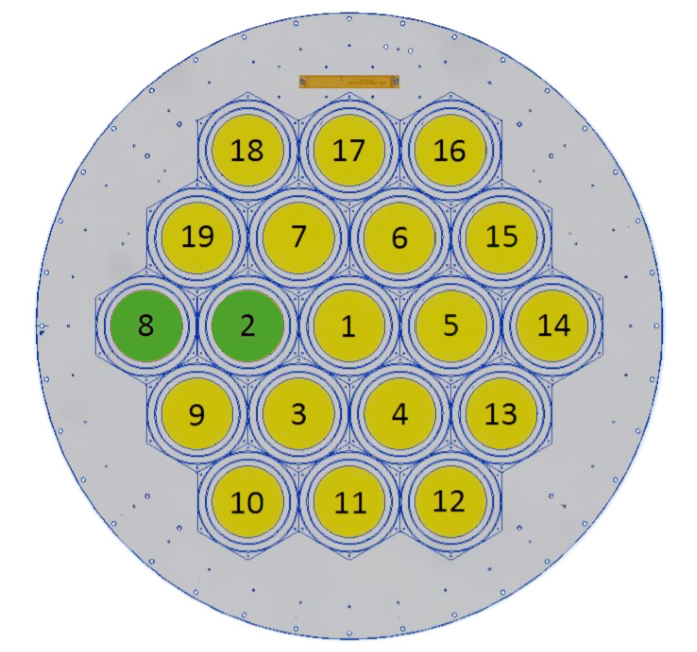}
	}
	\quad
	\subfigure[]{
		\includegraphics[width=0.29\linewidth]{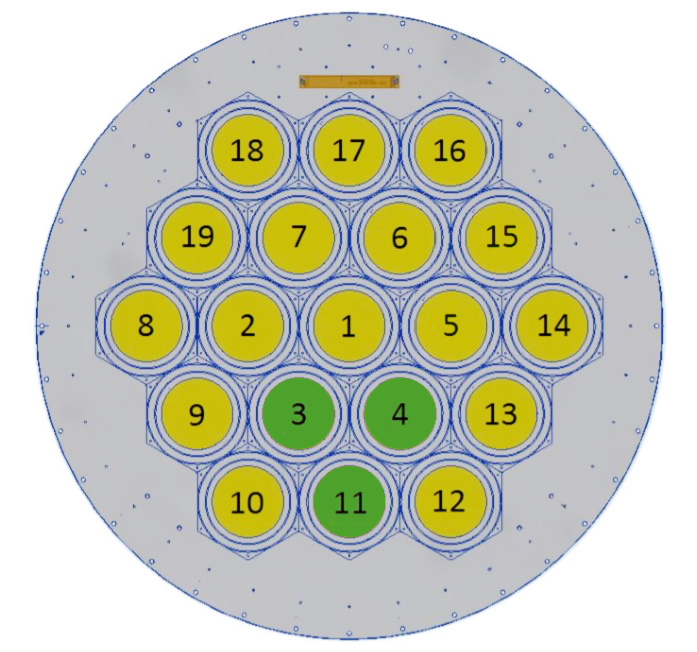}
	}
	\quad
	\subfigure[]{
		\includegraphics[width=0.29\linewidth]{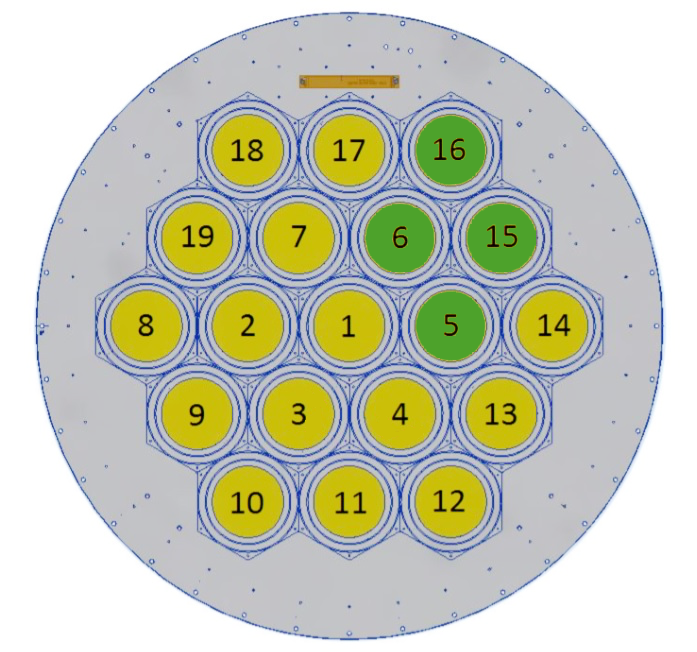}
	}
	\caption{Three examples of allowed signals of the MBCM blind search mode. This figure is same as the upper panels of Figure 1 in \citet{2023AJ....165..132L}.}
	\label{fig:beams}
\end{figure*}

\subsection{MBCM Blind Search Mode} \label{subsec:strategy}

Identifying ubiquitous man-made RFI remains a significant challenge in radio SETI research. As mentioned above, both MBCM targeted search mode and MBCM blind search mode are designed for the targeted SETI observations. However, the MBCM blind search mode is the most comprehensive and effective approach. It combines all 19 beams to search a larger sky area near the observation target (within the 19-beam-pointed sky, $\sim$ $11^{\prime}.6$) and can detect signals appearing between beams \citep{2023AJ....165..132L}. This mode offers advantages over the MBCM targeted search mode, which can only detect signals in the central beam. Therefore, we choose the MBCM blind search mode for this SETI observation.

The detailed criteria of the MBCM blind search mode are shown in \citet{2023AJ....165..132L}. There are four types of allowed beam coverage arrangement which may potentially come from the ETI for the MBCM blind search mode: (1) one beam, could be any of the 19 beams; (2) two adjacent beams, such as Beams 2 and 8 (Figure \ref{fig:beams}(a)); (3) three adjacent beams forming an equilateral triangle, such as Beams 3, 4, and 11 (Figure \ref{fig:beams}(b)); and (4) four adjacent beams forming a compact rhombus, such as Beams 5, 6, 15, and 16 (Figure \ref{fig:beams}(c)). However, since the radiation angle of RFI is typically large so that any other type of beam coverage arrangement, including coverage of non-adjacent beams, or coverage of more than four adjacent beams, or coverage of three or more beams on a line, are all rejected as RFI. We record the data of all 19 beams. Beam 1 (the central beam) keeps tracking the target, with all other beams serving as reference beams. Any detection with a signal-to-noise ratio (S/N) above the threshold is categorized as a $hit$. Among all $hits$, we refer to those that meet the MBCM blind search mode criteria as $events$; otherwise, they are categorized as RFIs and immediately rejected.

\subsection{Polarization Techniques} \label{subsec:strategy_polar}

Most targeted SETI research in recent years only focus on analyzing Stokes I data to detect narrowband signals \citep{2017AJ....153..110G,2020AJ....159...86P,2020AJ....160...29S,2021NatAs...5.1148S}. The observation data output by the FAST multibeam digital backend includes four polarization channels: $XX$, $YY$, $X^{\ast}Y$, $Y^{\ast}X$ \citep{2020RAA....20...64J}. We process data from two orthogonal linear polarization directions $XX$ and $YY$ respectively, thereby providing a more comprehensive assessment for RFI removal than that of the commonly analyzed Stokes I data. The above two polarization directions analysis have already been applied in previous studies \citep{2022AJ....164..160T,2023AJ....165..132L}. Specifically, in the same set of targeted SETI observations, we observe multiple targets continously and record data of all 19 beams. The comparison of different targets within the same set of observations can be used to remove RFI. \citet{2022AJ....164..160T} proposed that the same kind of instrumental noise would lead to the similar polarization characteristics, that is, a stronger or weaker of relative intensity in $XX$ than $YY$ of sevaral $events$ indicate their same origin. This is an effective tool to identify and remove instrumental interference.

\section{Data Analysis} \label{sec:analysis}

We record data across $1.0-1.5$ GHz on the SETI backend using the 19 beam receiver, with the number of frequency channels of the FAST SETI backend of 65536 * 1024, a frequency resolution of $\sim$ 7.5 Hz, and an integration time of $\sim$ 10 s for each spectrum \citep{fastmanref}. Each FITS file \citep{1981A&AS...44..363W} records data from four polarization channels ($XX$, $YY$, $X^{\ast}Y$, $Y^{\ast}X$) from two spectra of each beam. We process the data in two orthogonal linear polarization directions ($XX$ and $YY$) separately to show the polarization characteristics of the signals and identify RFIs more efficiently. The $hits$ in each beam are searched by the TurboSETI package \citep{2017ApJ...849..104E,2019ascl.soft06006E}, which could search for narrow-band signals in the two-dimensional time-frequency power spectrum using a tree search algorithm \citep{2013ApJ...767...94S}. We convert the FITS file obtained from the observations into a Filterbank file, which could be opened for subsequent processing by the Blimpy package \citep{2019JOSS....4.1554P}.

Two key parameters in the TurboSETI algorithm are the S/N threshold and the maximum drift rate (MDR), which make it possible to search for narrowband signals above an S/N threshold, as well as drift rates between $\pm$ MDR. Referring to previous SETI studies using TurboSETI \citep{2017ApJ...849..104E,2020AJ....159...86P,2020AJ....160...29S,2021AJ....162...33G,2021NatAs...5.1148S,2021AJ....161..286T}, we set the S/N threshold to 10. Due to the Doppler effect by the relative motion of the Earth and the target star, a narrowband signal transmitted from a extraterrestrial source drifts in frequency, and the drift rate is given by $\dot{\nu}=\nu_0 a / c$, where $\nu_0$ is the original frequency, $a$ is the relative acceleration between the transmitter and the receiver (the telescope), and $c$ is the speed of light. The MDR depends on the observing frequency band, since the drift rate is proportional to the transmission frequency of the signal. We set the MDR to 4 Hz $s^{-1}$, which is large enough for searching signals in the L-band \citep{2022ApJ...938....1L}. TurboSETI outputs the best-fit frequencies, drift rates, and S/Ns for $hits$ in all beams to a DAT file. In addition to using TurboSETI, we also use a de-drifting algorithm for some specific signals to calculate drift rates with higher accuracy \citep{2022AJ....164..160T}.

After acquiring all $hits$, 19 beams are used to search for $events$. Following \citet{2023AJ....165..132L}, we first list all possible valid masks based on the allowed beam coverage arrangement of the MBCM blind search mode criteria. This mask consists of 19 bits binary numbers, each representing a beam, with `0' and `1' representing `mask' and `allow', respectively. The signal collected by the receiver is also identified by a 19 bits binary number consisting of `0' and `1', known as the identification code. We set the frequency error range to $\pm$ $5 \delta v$ ($\delta v$ is the frequency resolution). It means that $hits$ within the frequency error range can be considered to have the same frequency and are marked as `1'; otherwise they are marked as `0'. If the number of `1's in the identification code is more than 4, we add an RFI mark directly. If the number of `1's is less than or equal to 4, we perform a `same-or operation' on the identification code and the list of valid masks (if the two values are the same, the result is 1, otherwise it is 0). In the above operation, if the result of the operation is 1, we add an ETI candidate mark and end this operation. If the result of the identification code with all valid masks is 0, we add an RFI mark and reject it.

The multibeam data processing in MBCM blind search mode outputs all $events$ for each target to a CSV file. Since the effective frequency band of FAST L-band receiver is $1.05-1.45$ GHz, the top and bottom 50 MHz are not in its design band \citep{2011IJMPD..20..989N,2020RAA....20...64J}, so we drop the 50 MHz wide $events$ at both ends.

\begin{figure*}[!htp]
	
	\centering
	\subfigure[]{
		\includegraphics[width=0.98\linewidth]{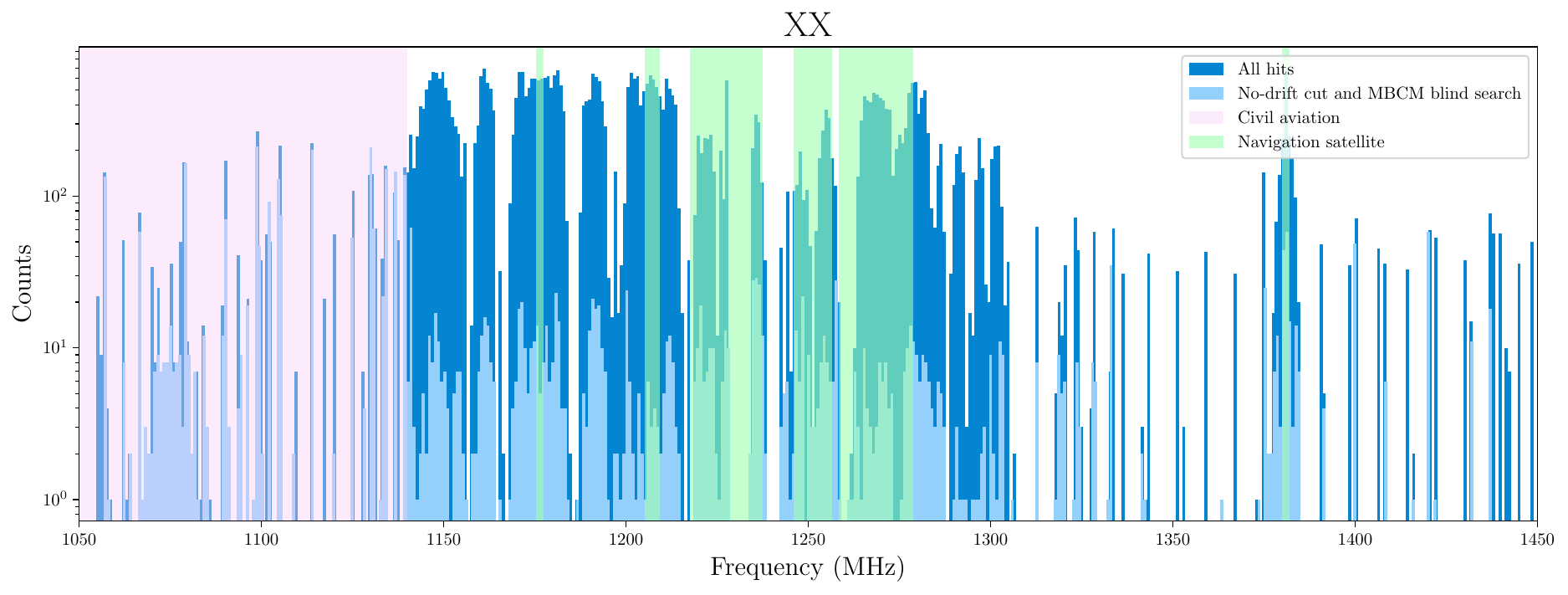}
	}
	
	\subfigure[]{
		\includegraphics[width=0.47\linewidth]{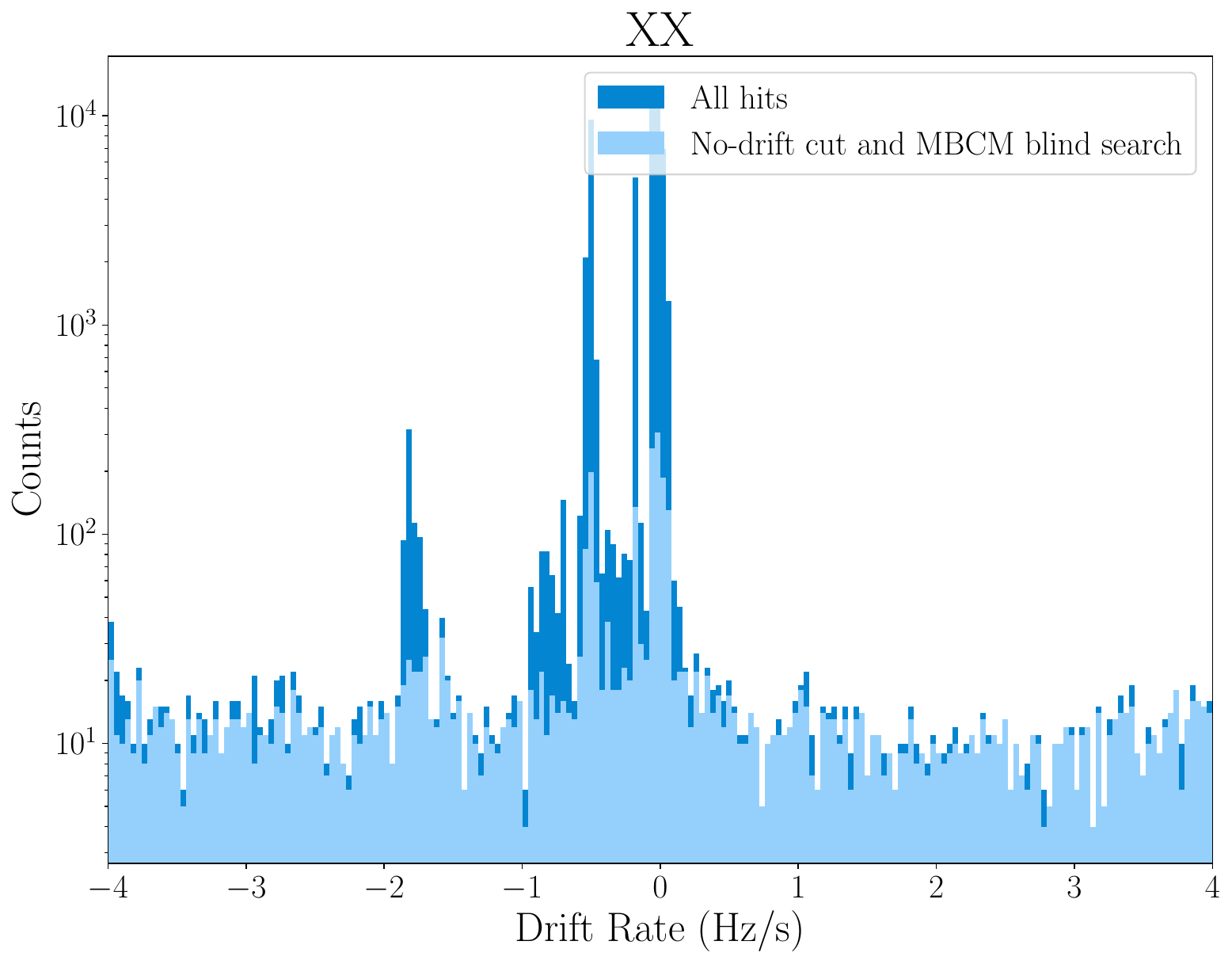}
	}\subfigure[]{
		\includegraphics[width=0.47\linewidth]{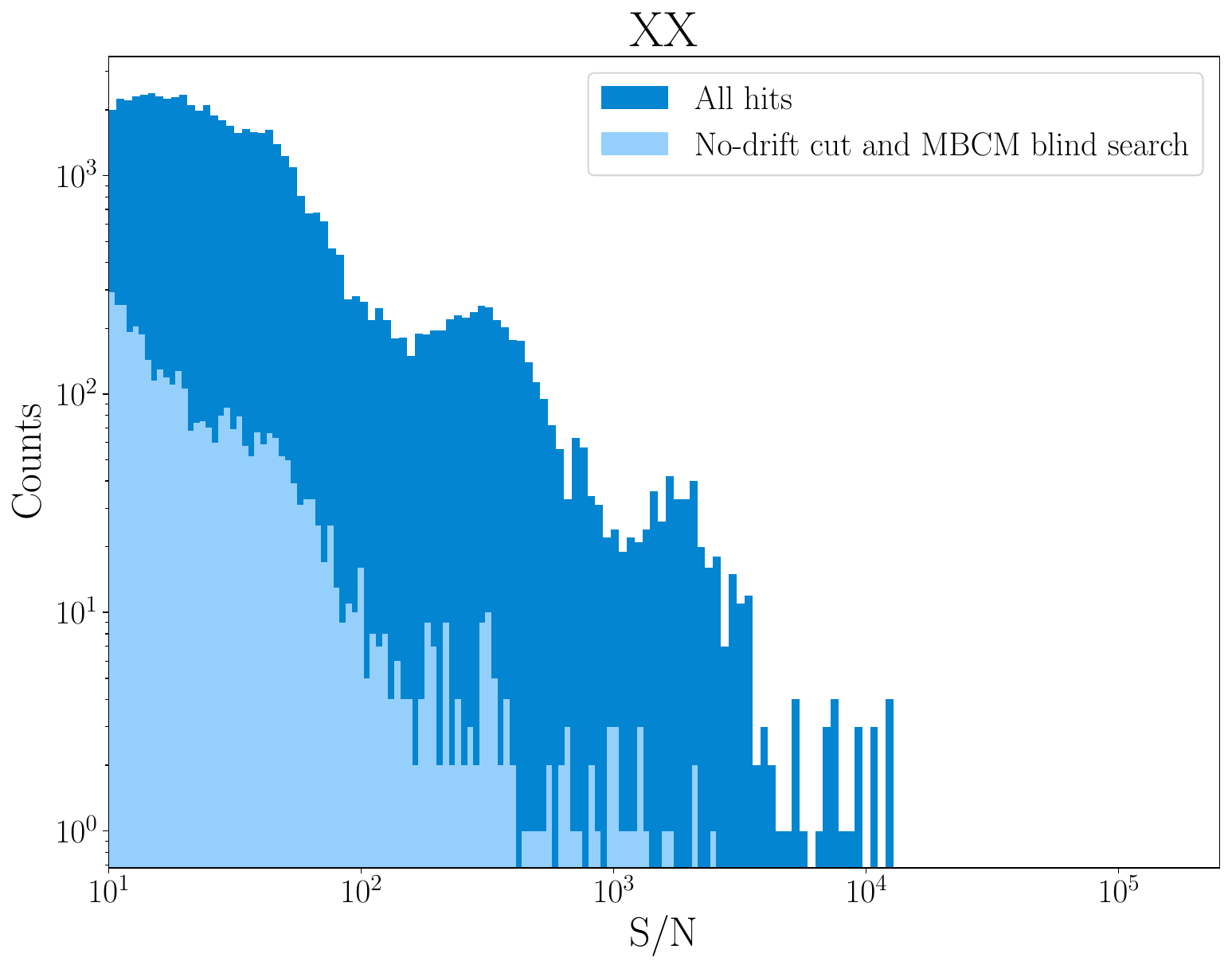}
	}
	\caption{The distributions of frequency, drift rate, and S/N in $XX$ polarization. The frequency bands of cataloged interference sources are displayed on the frequency panel. The dark blue bars indicate the distributions of the $hits$ detected by all 19 beams. The light blue bars indicate the distributions of the $events$ removed the zero drift rate signals and detected by the MBCM blind search mode.}
	\label{fig:statistcs_XX}
\end{figure*}

\begin{figure*}[!htp]
	
	\centering
	\subfigure[]{
		\includegraphics[width=0.98\linewidth]{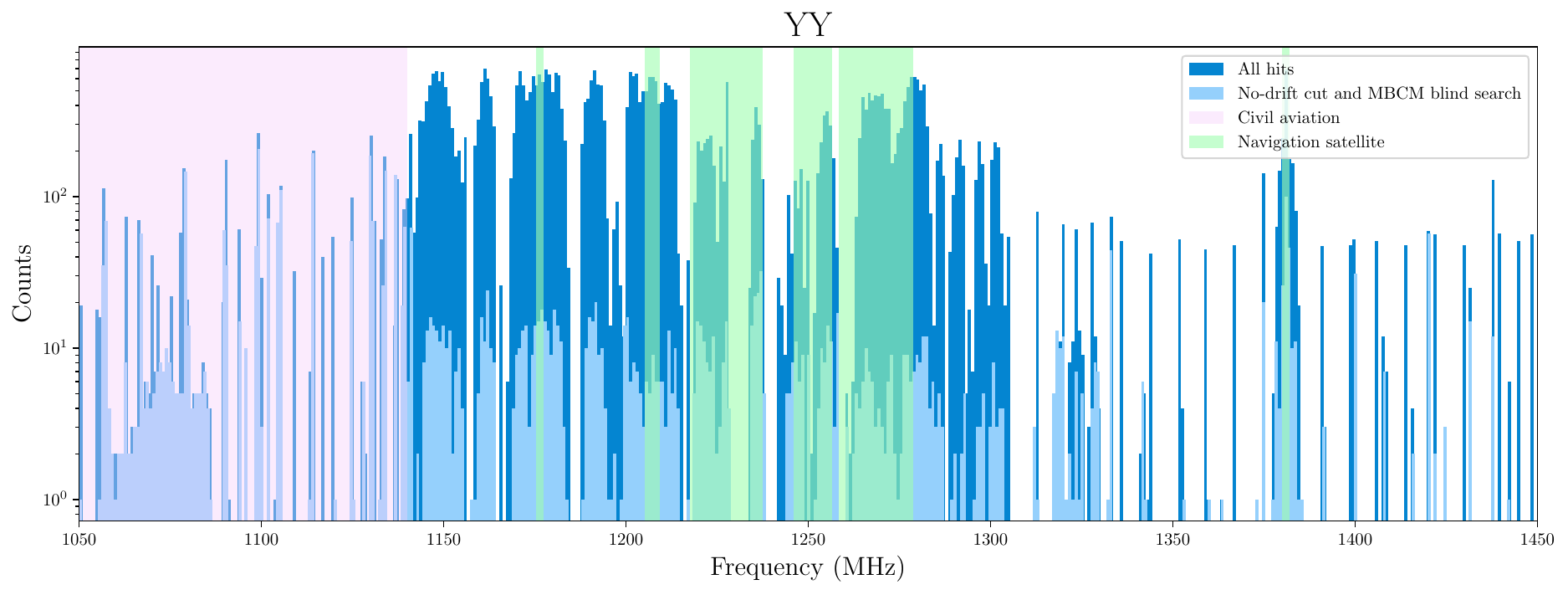}
	}
	
	\subfigure[]{
		\includegraphics[width=0.47\linewidth]{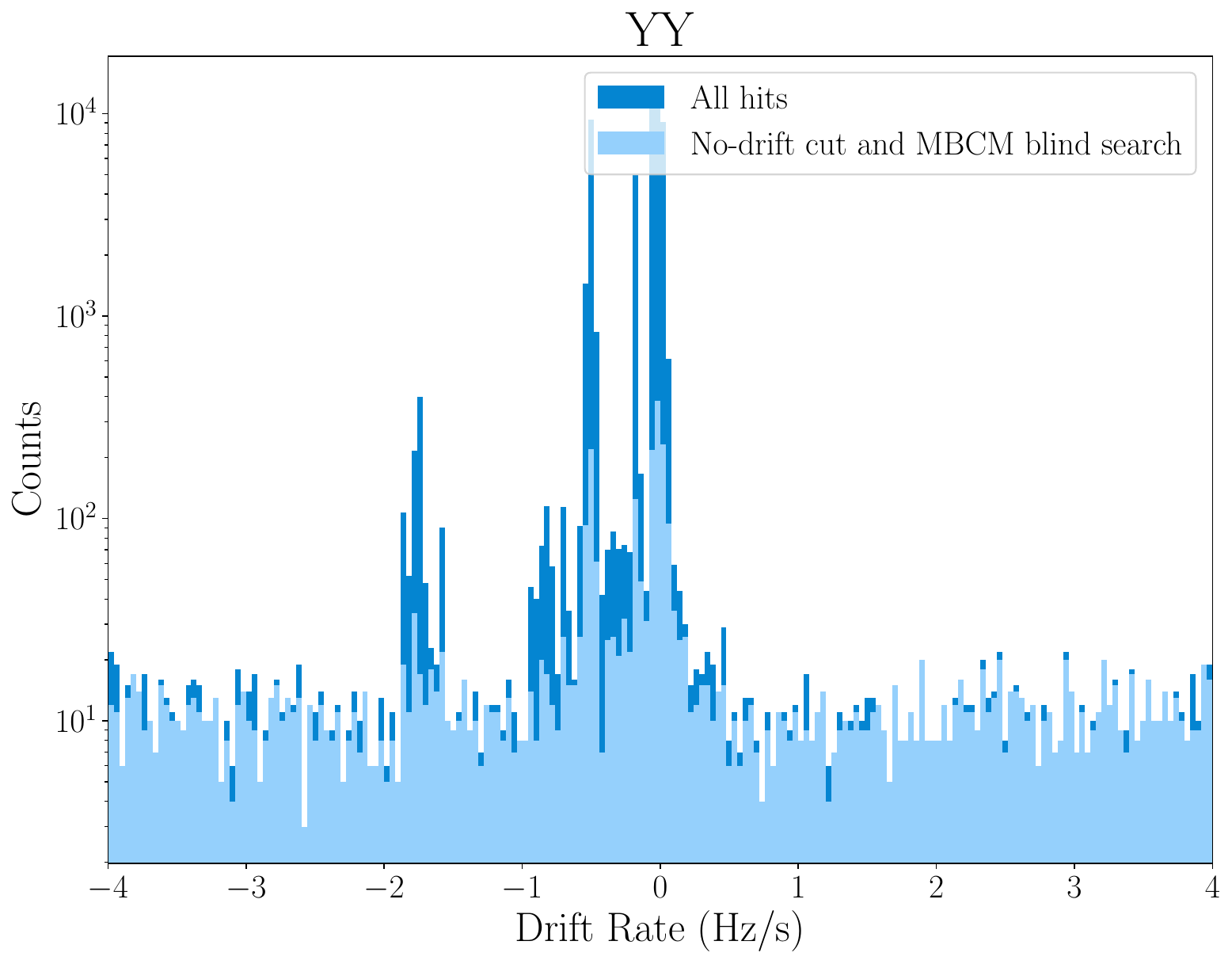}
	}\subfigure[]{
		\includegraphics[width=0.47\linewidth]{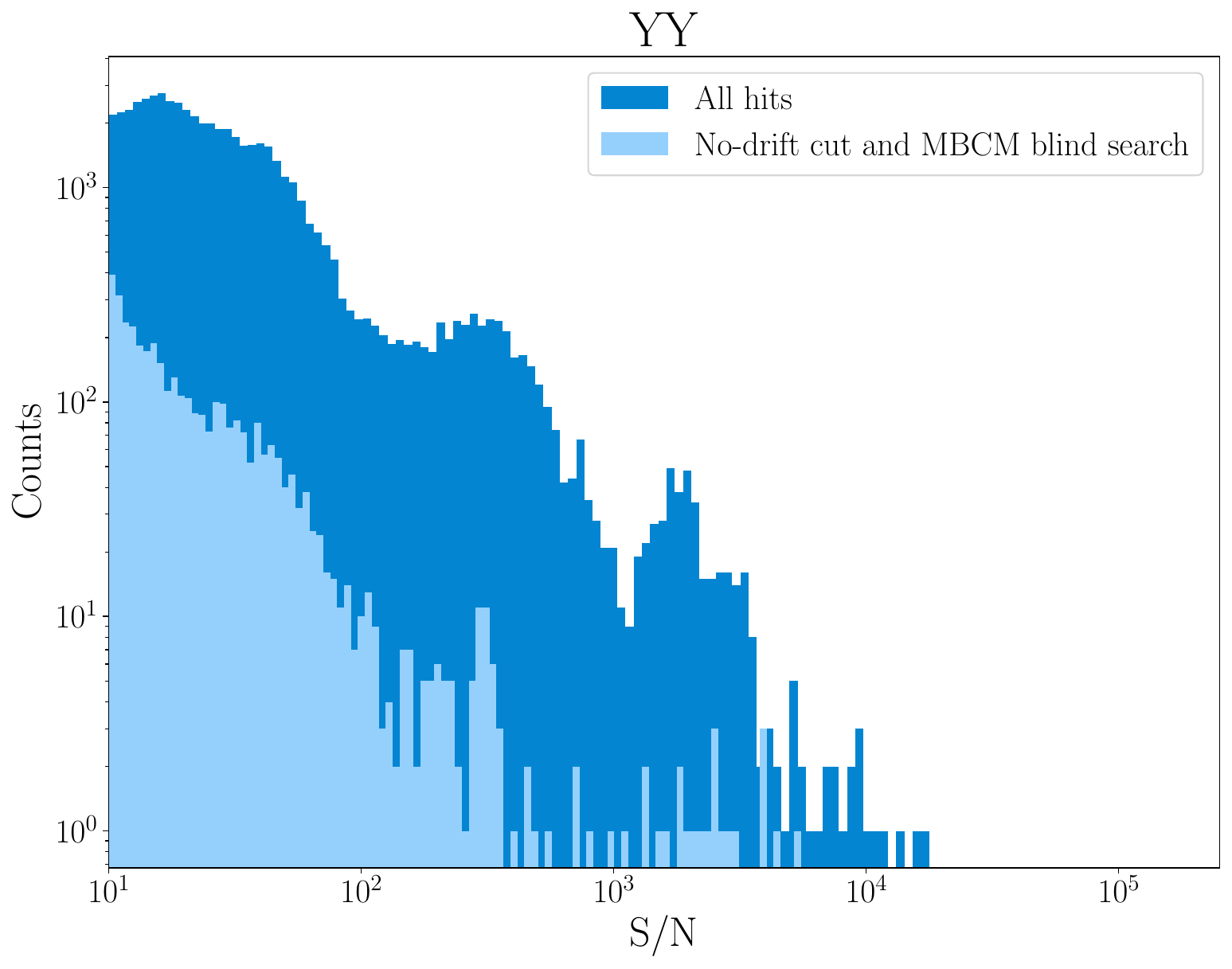}
	}
	\caption{The distributions of frequency, drift rate, and S/N in $YY$ polarization.}
	\label{fig:statistcs_YY}
\end{figure*}

\section{Results} \label{sec:results}

\subsection{Signal Statistics and RFI Removal} \label{subsec:statistics}

We select 55746 $hits$ from $XX$ and 55373 $hits$ from $YY$ in all 19 beams. First, we remove any $hits$ that have a drift rate of 0 in the geocentric coordinate system, since these signals are likely to come from ground RFI. Then we find 3787 $events$ of $XX$ and 3725 $events$ of $YY$ based on the MBCM blind search mode criteria in Section \ref{subsec:strategy}. The distribution of all $hits$ detected in all 19 beams and the $events$ detected by the MBCM blind search mode are shown in Figures \ref{fig:statistcs_XX} and \ref{fig:statistcs_YY}, respectively. In both figures, if an $event$ is detected by more than one beam, its parameters (e.g., frequency, drift rate, and S/N) are represented by the average of these beams. As can be seen in Figures \ref{fig:statistcs_XX} and \ref{fig:statistcs_YY}, there is no significant difference between the distributions of $XX$ and $YY$ polarizations, which suggests that most $hits$ and $events$ do not have a preference for linear polarization in $X$ and $Y$.

According to the RFI environmental monitoring at the FAST site, there are two sources of RFI in the observation band: civil aviation and navigation satellites \citep{2021RAA....21...18W}. In the $XX$ polarization, there are 2164 $\bm{events}$ located in the civil aviation band (1030-1140 MHz) and 548 $\bm{events}$ located in the navigation satellite bands (1176.45 $\pm$ 1.023 MHz, 1207.14 $\pm$ 2.046 MHz, 1227.6 $\pm$ 10 MHz, 1246.0-1256.5 MHz, 1268.52 $\pm$ 10.23 MHz, and 1381.05 $\pm$ 1.023 MHz). In the $YY$ polarization, there are 1995 $\bm{events}$ located in the civil aviation band and 616 $\bm{events}$ located in the navigation satellite bands. Apart from these, the instruments in the backend  can cause significant interference, i.e., instrumental RFI \citep{2022AJ....164..160T}. The mechanisms causing RFI can be extremely complex. To put it simply, instrumental RFIs are generated at specific frequencies that can be obtained by linear combinations of the nominal frequencies \footnote{\url{https://github.com/ska-sa/roach2_hardware/blob/master/release/rev2/A/BOM/ROACH-2_REV2_BOM.csv}}  (33.3333, 125.00, and 156.25 MHz, etc.) of the crystal oscillators on the Roach-2 FPGA board at the instrument backend, so these signals are essentially intermodulation products. We directly remove the $events$ near these specific frequencies, e.g. $\sim$ 1066.66, $\sim$ 1333.34, $\sim$ 1400.01, $\sim$ 1124.99 MHz, etc. The number of instrumental RFIs includes 319 of $XX$ and 320 of $YY$, which account for approximately 8\% of all $events$.

\begin{figure*}
	\centering
	\includegraphics[width=0.9\linewidth]{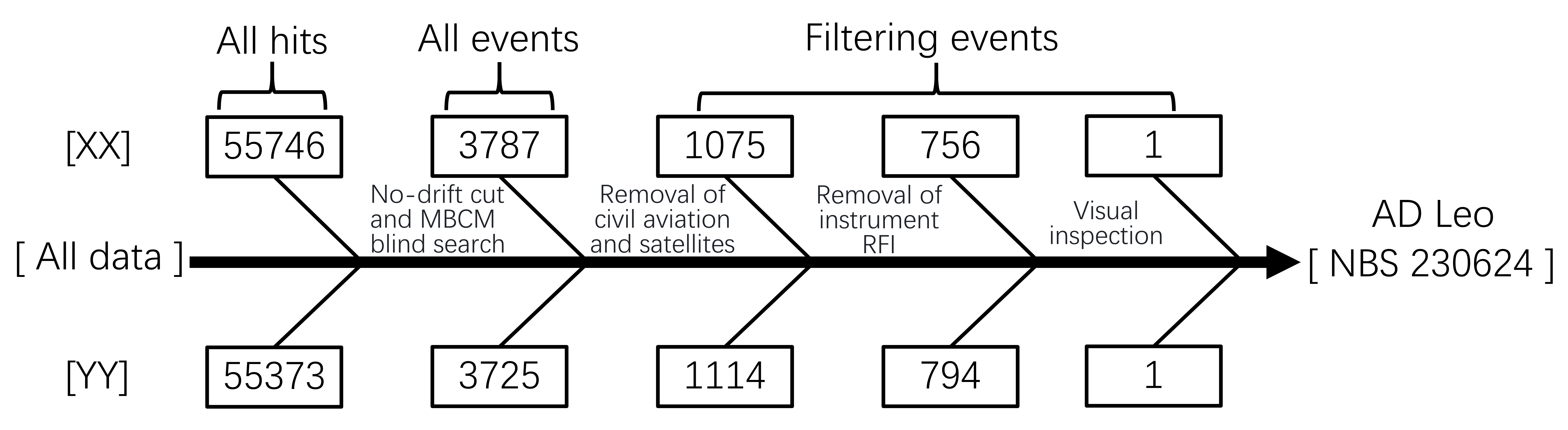}
	\caption{A flow chart of the search pipeline and results. The pipeline consists of five steps. All $hits$ in the 19 beams are obtained in the first step. All $events$ are obtained using the MBCM blind search mode in the second step. The last three steps use RFI removal methods to filter the $events$. The numbers in each step are shown in the boxes, where the first row is for $XX$ and the second row is for $YY$. Finally, NBS 230624 is found in both $XX$ and $YY$ polarizations.}
	\label{fig:pipeline}
\end{figure*}

Finally, we re-examine all remaining $events$ by visual inspection of the dynamic time spectrum (waterfall plot) to find the most particular $events$ for detailed analysis. We use the Blimpy package \citep{2019JOSS....4.1554P} to plot waterfall plots of $events$ for 19 beams. We find that most $events$ are obvious false positives, for example: (1) A narrowband signal is present in a beam, but TurboSETI do not identify it as a $hit$. Because it is weak and do not reach the S/N threshold; (2) No narrowband signal is present in the beam, but TurboSETI incorrectly identifies it as a $hit$. For the latter case, we can de-drift and time-integrate the spectrum of that beam to prove there is indeed no signal. The vast majority of the remaining $events$ are false positives. After visual inspection, only one $event$ remained, which is detected in Beam 2 during the AD Leo observations (hereafter NBS 230624, where NBS means narrowband signal, and 230624 is the date we observed it). The flow chart of search pipeline and results are shown in Figure \ref{fig:pipeline}.

\begin{figure*}
	\centering
	\includegraphics[width=0.85\linewidth]{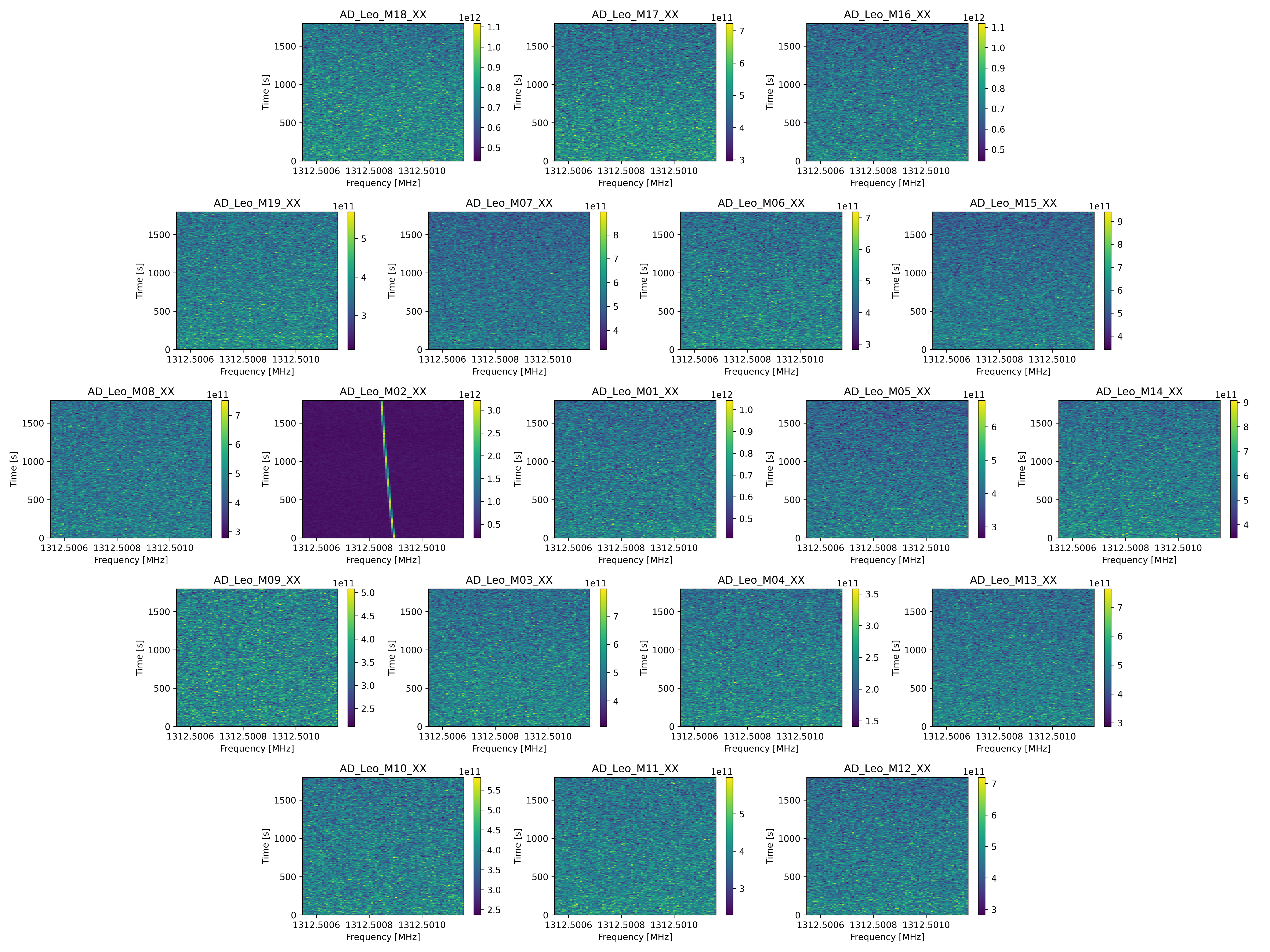}
	\caption{An $event$ detected at $\sim$ 1312.50 MHz toward AD Leo (NBS 230624). This figure shows $XX$ polarization of all 19 beams.}
	\label{fig:AD_Leo_XX_19}
\end{figure*}

\subsection{Analysis on NBS 230624} \label{subsec:most special}

\begin{figure*}
	\centering
	\includegraphics[width=0.98\linewidth]{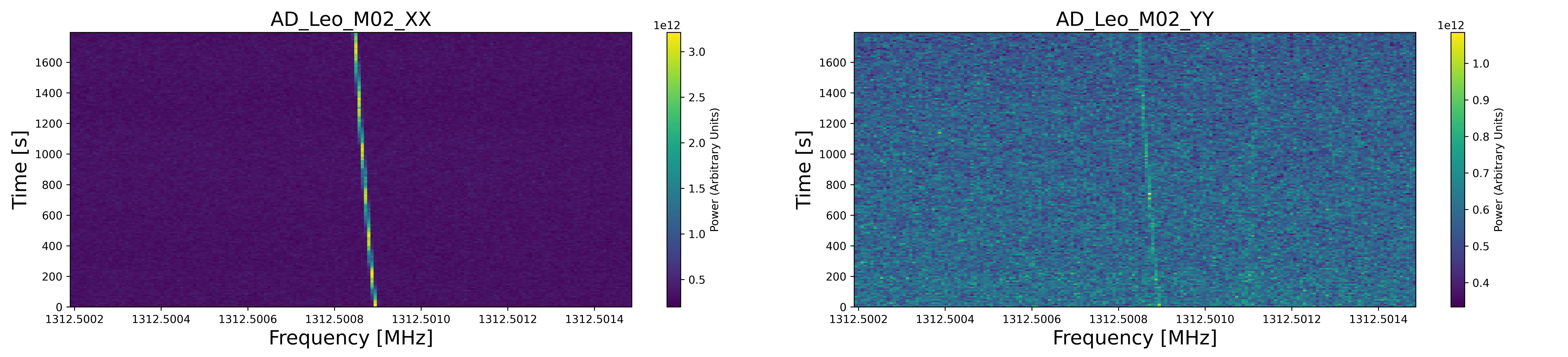}
	\caption{Waterfall plots of AD Leo NBS 230624 in Beam 2. The left panels show $XX$, and the right panels show $YY$.}
	\label{fig:1312}
\end{figure*}

NBS 230624 is detected at $\sim$ 1312.50 MHz during the observation of AD Leo, with a drift rate of about $-$0.0249 Hz $s^{-1}$. Its frequency cannot be obtained from a linear combination of the nominal frequency of the crystal oscillators, and its drift rate is within the range of a transmitter moving with an exoplanet. But it is not present in the central beam, only present in Beam 2. For further analysis, we use the de-drifting algorithm for all 19 beams in the $XX$ and $YY$ polarizations, respectively. The S/Ns of Beam 2 in $XX$ and $YY$ polarizations are quite different, with $XX$ (S/N $\sim$ 596.5516) being much stronger than $YY$ (S/N $\sim$ 21.03254). No signals are detected in any other beams. The waterfall plots of NBS 230624 of all 19 beams in $XX$ polarization are shown in Figure \ref{fig:AD_Leo_XX_19}, and those of Beam 2 in both $XX$ and $YY$ polarizations are presented in Figure \ref{fig:1312}. More detailed information on NBS 230624 obtained by the de-drifting algorithm is shown in Table \ref{table:special}.

\begin{figure*}
	\centering
	\includegraphics[width=0.98\linewidth]{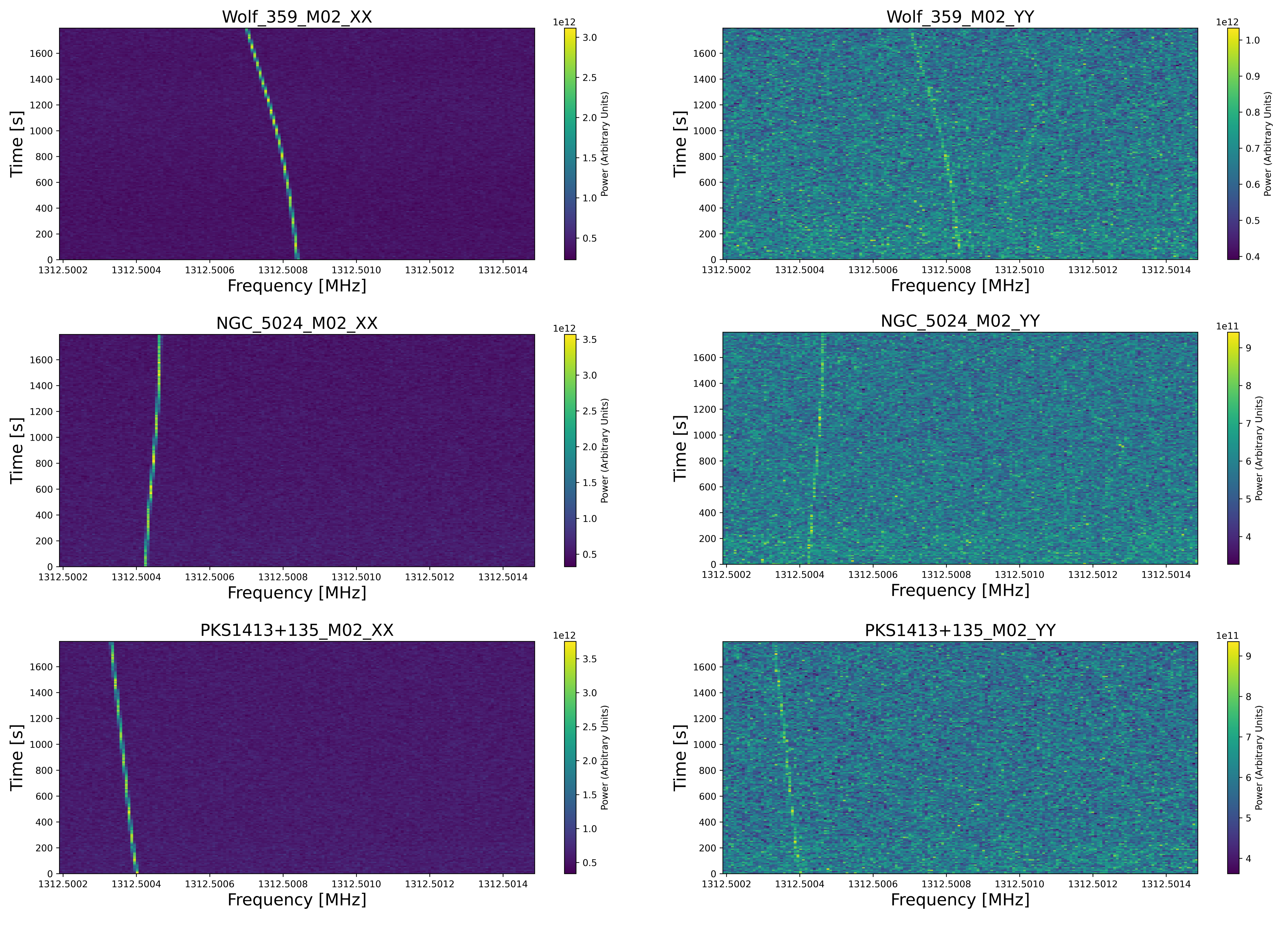}
	\caption{Waterfall plots around 1312.50 MHz of the observations toward other three sources. The left panels show $XX$, and the right panels show $YY$. They are only present in Beam 2 and appear obviously stronger in $XX$ than $YY$, consistent with NBS 230624.}
	\label{fig:others}
\end{figure*}

According to Section \ref{subsec:strategy_polar}, instrumental RFIs from the same source have similar polarization characteristics. Therefore, we visually re-examine the waterfall plots of all 19 beams at 1312.50 MHz $\pm$ 2 kHz in $XX$ and $YY$ polarizations of the other two targets, as well as the data for AD Leo in other frequency channels. We discover a narrowband signal in Beam 2 during the Wolf 359 observation (see the upper panals of Figure \ref{fig:others}). Its starting frequency (1312.50087 MHz) and polarization characteristic (stronger in $XX$ than $YY$) are similar to those of NBS 230624. Due to its rather special drift behavior, it is not captured initially by our search pipeline. In addition, we also examine the data at 1312.50 MHz $\pm$ 2 kHz of other types of targets observations on June 24, 2023\footnote{On June 24, 2023, besides observing the three nearby M dwarfs mentioned above, we also observed globular clusters and radio galaxies. The results will shown in our future studies.}. Two other signals during the observations toward NGC 5024 and PKS 1413+135\footnote{NGC 5024 ($13^{\mathrm{h}}\ 12^{\mathrm{m}}\ 55.25^{\mathrm{s}}$, $+18^{\circ}\ 10^{\prime}\ 05.4^{\prime\prime}$) is a globular cluster and PKS 1413+135 ($14^{\mathrm{h}}\ 15^{\mathrm{m}}\ 58.82^{\mathrm{s}}$, $+13^{\circ}\ 20^{\prime}\ 23.7^{\prime\prime}$) is a radio galaxy. Their observations were conducted using the same strategy as this work. The start MJDs are 60119.428472 and 60119.484027 for NGC 5024 and PKS 1413+135, respectively.} are found (see the middle and lower panals of Figure \ref{fig:others}). Although the frequencies of the two signals are slightly lower than that of NBS 230624, and their drift behaviors are different, we find that they are much stronger in $XX$ than $YY$, and only visible in Beam 2, which is highly consistent with the characteristics of NBS 230624. In a word, we find other three signals with the following similar characteristics to NBS 230624: (1) the same beam coverage (only in Beam 2); (2) similar starting frequency (in the range of $\pm$ 0.5 kHz.); and (3) similar polarization characteristic (stronger in $XX$ than $YY$). Based on the above evidence, we conclude that NBS 230624 is most likely to come from the same instrumental source with these three signals, but their exact origin is still unknown.

\begin{figure*}
	\centering
	\includegraphics[width=0.98\linewidth]{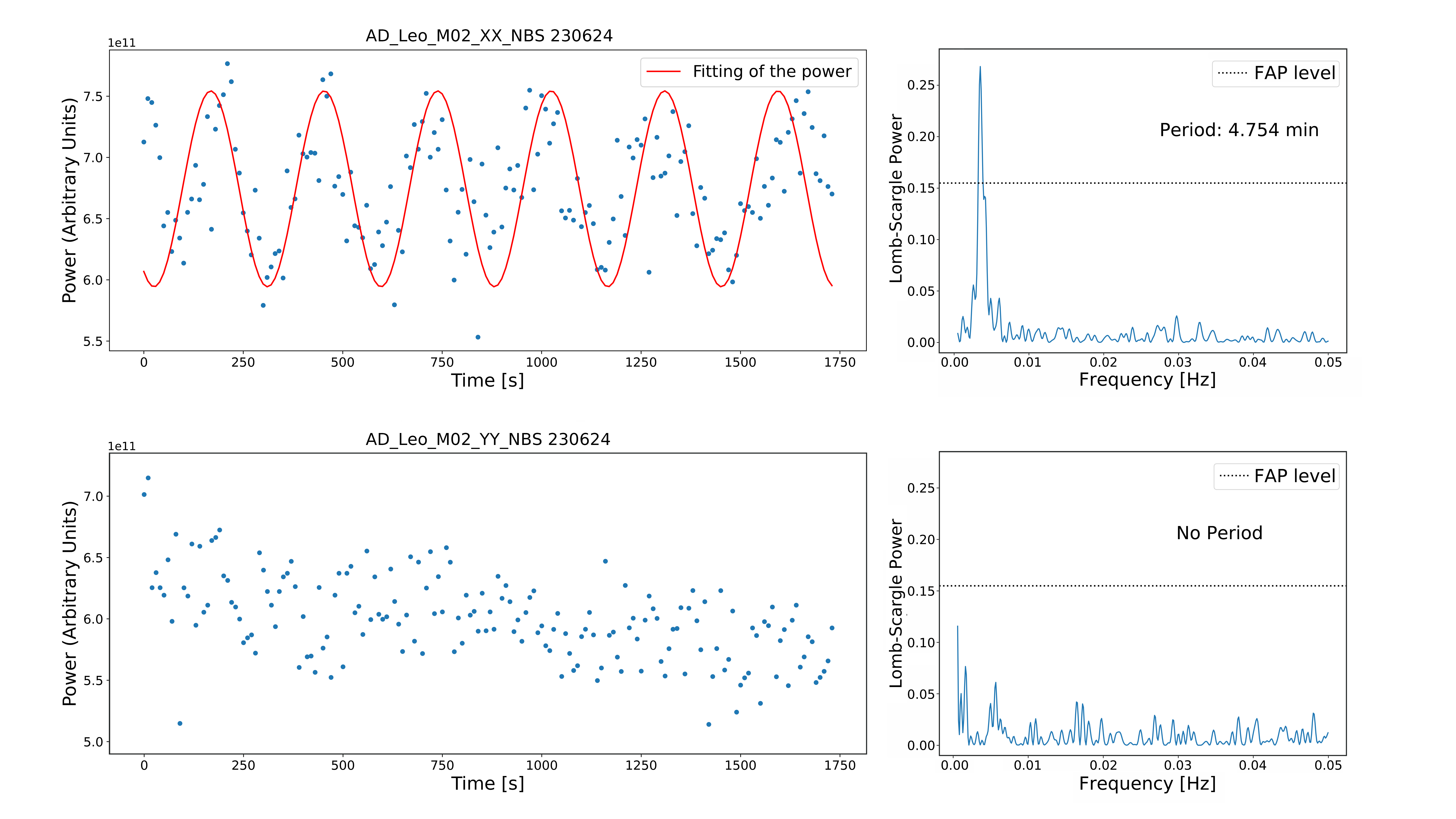}
	\caption{The power variation of NBS 230624 over time and the Lomb-Scargle periodogram. The left panels show the power variation of the signal bins of NBS 230624 over time for $XX$ and $YY$ polarizations. The right panels are the Lomb-Scargle periodograms accordingly. The solid line in the upper left panel denote a first-order Fourier function fitting, while the dotted black lines in the right panels are the FAP level at 0.001. See the main text for details.}
	\label{fig:fft}
\end{figure*}

\begin{table*}[!ht]
	\caption{Detailed Information of AD Leo NBS 230624}
	\centering
	\begin{tabular}{ccccccc}
		\hline \hline Beam No. & Polarization & Start Frequency & Start MJD & Drift Rate & S/N\\
		\colhead{} & \colhead{} & (MHz)  & \colhead{} & (Hz $s^{-1})$ & \colhead{}\\
		\hline M02 & $XX$ & $1312.50088$ & $60119.290972$ & $-0.0249$ & $596.5516$\\
		M02 & $YY$ & $1312.50088$ & $60119.290972$ & $-0.0249$ & $21.03254$\\
		\hline
	\end{tabular}
	\label{table:special}
\end{table*}

\section{Discussions} \label{sec:discussions}

\subsection{Sensitivity} \label{subsec:EIRP}

The sensitivity of a radio SETI observation is mainly determined by the effective collecting area and the system noise  of the telescope, and the performance of a telescope can be measured by the system equivalent flux density (SEFD):
\begin{equation}
   \mathrm{SEFD}=\frac{2 k_{\mathrm{B}} T_{\mathrm{sys}}}{A_{\mathrm{eff}}},
\end{equation}
where $k_B$ is the Boltzmann constant, $T_{sys}$ is the system temperature, and $A_{\mathrm{eff}}$ is the effective collecting area ($A_{\mathrm{eff}}=\eta A$, where $A$ is the physical collection area and $\eta$ is an efficiency factor between 0 and 1). The sensitivity of the radio telescope is proportional to the ratio $A_{eff}/T_{sys}$ \citep{2016era..book.....C,2017isra.book.....T}. The sensitivity of the FAST L-band 19 beam receiver is about 2000 $m^2/K$ \citep{2011IJMPD..20..989N,2016RaSc...51.1060L,2020RAA....20...64J}.

For observations searching for narrowband signals, the signal bandwidth is generally narrower than the frequency resolution. In this case, the minimum detectable flux $S_{min}$ is:
\begin{equation}
	S_{\text {min}}=\mathrm{S} / \mathrm{N}_{\text {min }} \mathrm{SEFD} \sqrt{\frac{\delta \nu}{n_{\mathrm{pol}} \tau_{\mathrm{obs}}}},
\end{equation}
where $\mathrm{S} / \mathrm{N}_\mathrm{min}$ is the S/N threshold, $\delta \nu$ is the frequency resolution, $n_\mathrm{pol}$ is the polarization number, and $\tau_\mathrm{obs}$ is the observation duration \citep{2017ApJ...849..104E}. Since we deal with the data of $XX$ and $YY$ separately, the polarization number is 1. It is worth noting that this expression is different from the $S_{min}$ for observations of astrophysical signals. The bandwidth of the astrophysical signals is much wider than the frequency resolution, so the unit of $S_{min}$ is Jy (1 Jy = $10^{-26}$ $\mathrm{W}$ $\mathrm{m}^{-2}$ $\mathrm{Hz}^{-1}$). For observations of narrowband signals, however, the unit of $S_{min}$ is $\mathrm{W} / \mathrm{m}^2$, not Jy. With the S/N threshold of 10, the frequency resolution of about 7.5 Hz, and the observation duration of 1800 s, we calculate the $S_{\min}=8.91 \times 10^{-27}$ $\mathrm{W} / \mathrm{m}^2$ (0.891 Jy Hz) in our observations.

Assuming that the transmitter radiates an isotropic ETI signal, we need to consider the distance of the observation sources when we discuss the sensitivity of targeted SETI observations. The minimum detectable equivalent isotropic radiated power ($\operatorname{EIRP}_{\min}$) is defined as:
\begin{equation}
	\operatorname{EIRP}_{\min}=4 \pi d^2 S_{\min },
\end{equation}
where $d$ is the distance of the target. For the three M dwarfs in our observations, the $\operatorname{EIRP}_{\min}$ that can be achieved for Wolf 359, AD Leo, and TVLM 513-46546 are $6.19 \times 10^{8}$ W, $2.63 \times 10^{9}$ W, and $1.23 \times 10^{10}$ W, respectively, which achieves a very high sensitivity.

Depending on the ability to use energy, \citet{1964SvA.....8..217K} proposed a classification scheme for technological civilizations. A Kardashev Type I civilization is defined as one that is able to harness all the stellar energy falling on their planet, e.g., about $10^{17}$ W for an Earth-like planet around a Sun-like star. A Kardashev Type II civilization is defined as one that can harness all the energy produced by its star, e.g., about $10^{26}$ W for the Sun-like star. A Kardashev Type III civilization would be able to utilize all the energy produced by all the stars in a galaxy, e.g., about $10^{36}$ W for a Milky Way like galaxy. Thus, our FAST SETI survey of the three targets will be able to identify any signals from Type I, II and III civilizations. Compared to the Arecibo planetary radar of about $10^{13}$ W EIRP, the weakest signal we can detect is well within the reach of current human technology.

\subsection{The Power Variation in Two Polarizations}
\label{subsec:polarization}

We notice that NBS 230624 and the other three signals (Figure \ref{fig:others}) show obvious fluctuations in power over time. Thus, we analyse their variation properties in both polarizations. For the NBS 230624, we first extract the data of the signal and three frequency channels on either side of it (a total of 7 channels) as the signal bins. Then we calculate the average power of the 7 channels at each sampling time in $XX$ and $YY$ polarizations, respectively. The power variations of the signal bins over time in the $XX$ and $YY$ polarizations are displayed in the left panels of Figure \ref{fig:fft}. As seen in the panels, the power variation of the $XX$ polarization is periodic, while that of the $YY$ polarization is non-periodic. In order to search for its periodicity, a Fourier analysis with Lomb-Scargle algorithm \footnote{https://docs.astropy.org/en/latest/timeseries/lombscargle.html} \citep{2018ApJS..236...16V} is conducted to the power variations in the $XX$ and $YY$ polarizations, as shown in the right panels of Figure \ref{fig:fft}. The most prominent peak in the upper right panel of Figure \ref{fig:fft} denote the period in the $XX$ polarization, i.e., 4.754 min. This is reinforced by the False Alarm Probability (FAP) level of 0.001, which means that there is 99.9\% certainty that this is a true period. The first-order Fourier function is fitted to the period mentioned above and shown in solid red line in the upper left panel of Figure \ref{fig:fft} to visually check its periodicity. For the $YY$ polarization, however, no significant peak is shown, leading to a non-detection of period. We also analyse the other three signals using the Lomb-Scargle algorithm, and find that they all exhibit no periodicity in either of the two polarizations. Regarding the NBS 230624 4.754-minute period of the power, we are unable to determine its source. It may come from man-made signals during the observations, or the instrument issues, etc. This periodicity could serve as a new criterion for RFI removal along with the development of observation, which will be detailed in our future study.

\section{Conclusions} \label{sec:conclusions}

We conduct searches for narrowband drifting radio signals toward three nearby M dwarfs with the FAST 19-beam receiver. Using the MBCM blind search mode, we record the data on the SETI backend with a frequency resolution of $\sim$ 7.5 Hz, and an integration time of ~10 s for each spectrum. We search the signals across $1.05–1.45$ GHz with drift rates within $\pm$ 4 Hz $s^{-1}$ and S/Ns above 10, in two orthogonal linear polarization directions separately.

The vast majority of detected $events$ are obvious RFIs, either false positives or known RFIs (civil aviation, navigation satellites, and intermodulation products caused by crystal oscillators). The most particular $event$, NBS 230624, which initially pique our interest. However, we eliminate the possibility of its extraterrestrial origin. In conclusion, our observations find no solid evidence for radio transmitters emitting between 1.05 and 1.45 GHz with an EIRP above $1.23 \times 10^{10}$ W for all three sources.

~\\
We sincerely appreciate the referee's suggestions, which helped us greatly improve our manuscript. We also sincerely thank Jian-Kang Li for the kind and useful discussions. This work was supported by National Key R$\&$D Program of China, No.2024YFA1611804, National SKA Program of China, No.2022SKA0110202 and China Manned Space Program through its Space Application System. This work was finished on the servers from FAST Data Center in Dezhou University.

\bibliography{arXiv-0227}{}
\bibliographystyle{aasjournal}

\end{document}